\documentclass[preprint,12pt]{elsarticle}

\usepackage{amssymb}
\usepackage{color}
\journal{Physics Letters B}

\newcommand{\Fig}[1]{Figure~\ref{#1}}

\begin{document}

\begin{frontmatter}



\title{
%     A new method for a lattice determination of NN potentials and
%     its application to the NN scattering length\\
%Beating noises for multi-baryon systems \\
%-- Extraction of baryonic potentials without ground state saturation in lattice QCD -- 
  Hadron-Hadron Interactions  from Imaginary-time Nambu-Bethe-Salpeter
  Wave Function on the Lattice}


\author[aics]{Noriyoshi Ishii}
\author[utsukuba,ccs]{Sinya Aoki}
\author[riken]{Takumi Doi}

\author[riken,Hongo]{Tetsuo~Hatsuda}
\author[Titech]{Yoichi~Ikeda}
\author[Fujisawa]{Takashi~Inoue}
\author[riken]{Keiko~Murano}
\author[ccs]{Hidekatsu~Nemura}
\author[ccs]{Kenji~Sasaki}

\author{
%\address{
(HAL QCD Collaboration)
\\
\bigskip
\includegraphics[width=0.35\textwidth]{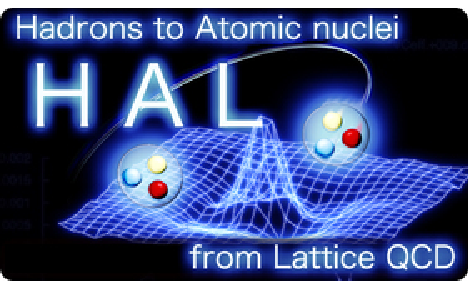}
}

\address[aics]{Kobe branch,
Center for Computational Sciences, University of Tsukuba,\\
in RIKEN Advanced Institute for Computational Science (AICS),\\
Portisland, Kobe 650--0047, Japan
%7--1--26 Portisland South, Kobe 650--0047, Japan
}
\address[utsukuba]{Graduate School of Pure and Applied Physics, University of Tsukuba, Tsukuba, \\ Ibaraki 305-8571, Japan}
\address[ccs]{Center for Computational Sciences, University of Tsukuba,Tsukuba, \\ Ibaraki 305-8577, Japan}
\address[riken]{Theoretical Research Division, Nishina Center, RIKEN, Wako 351-0198, Japan}
\address[Hongo]{Department of Physics, The University of Tokyo, Tokyo 113-0033, Japan}
%\address[IPMU]{IPMU, The University of Tokyo, Kashiwa 277-8583, Japan}
\address[Titech]{Department of Physics, Tokyo Institute of Technology, Meguro, Tokyo 152-8551, Japan}
\address[Fujisawa]{Nihon University, College of Bioresource Sciences, Fujisawa 252-0880, Japan}

\begin{abstract}
  Imaginary-time   Nambu-Bethe-Salpeter   (NBS)   wave   function   is
  introduced  to  extend   our  previous  approach  for  hadron-hadron
  interactions  on the  lattice.   Scattering states  of hadrons  with
  different energies encoded in  the NBS wave-function are utilized to
  extract  non-local  hadron-hadron  potential.   ``The  ground  state
  saturation'', which is  commonly used in lattice QCD  but is hard to
  be achieved for multi-baryons, is not required.  We demonstrate that
  the  present  method   works  efficiently  for  the  nucleon-nucleon
  interaction  (the potential  and  the phase  shift)  in the  $^1S_0$
  channel.
\end{abstract}


\end{frontmatter}

% \linenumbers


\newcommand{\Ref}[1]{Ref.~\cite{#1}}
\newcommand{\Eq}[1]{Eq.(\ref{#1})}
\newcommand{\Sect}[1]{Sect.~\ref{#1}}
\newcommand{\agt}{\raisebox{-.4ex}{\rlap{$\sim$}} \raisebox{.4ex}{$>$}}

\section{Introduction}

Euclidean  correlation functions are  dominated by  contributions from
the  corresponding  lowest-energy states  at  sufficiently large  time
separation $t$. This property,  called the ground state saturation, is
heavily used  in lattice QCD  to extract various  hadronic observables
such as masses, decay constants and other matrix elements.  The ground
state   saturation,  however,   is  difficult   to  be   achieved  for
multi-baryon systems.  For example,  the signal-to-noise ratio for the
correlation of $n$-nucleons reads ~\cite{Lepage:1989hd}:
\begin{equation}
\left( \frac{\cal S}{\cal N} \right)_n \sim  e^{-  n ( m_N - 3m_{\pi}/2) t},
\label{eq:S-N}
\end{equation}
where  $m_N$ and  $m_\pi$  are the  nucleon  mass and  the pion  mass,
respectively.  This relative enhancement of statistical noise at large
$t$  for  $m_N -  3m_{\pi}/2  >0$ is  a  common  problem for  baryonic
systems, even for a single baryon $(n=1)$.

In  addition, there exists  another problem  for the  multi-hadrons at
large $t$ caused by the small splitting between the ground and the 1st
excited states for large  volume.  For example, in the nucleon-nucleon
($NN$) system, the minimum splitting is estimated as
\begin{equation}
\Delta E \simeq \frac{{\bf p}_{\rm min.}^2}{m_N} = \frac{1} {m_N} \frac{ (2\pi)^2}{L^2},
\end{equation}
where $L$  is a spatial extension  of the lattice.  If  $L\simeq 6$ fm
and $m_N  \simeq 1$  GeV, we have  $ \Delta  E \simeq 43$  MeV $\simeq
1/(4.6 \,  {\rm fm})$.  The ground state  saturation requires $  t \gg
(\Delta E)^{-1} \simeq 4.6$ fm, which corresponds to $t/a \gg 46 $ for
the lattice spacing $a\simeq 0.1$ fm.  It is very difficult to extract
signals at such large $t$ due to the bad behavior of statistical noise
in   Eq.~(\ref{eq:S-N}).   To  avoid   these  problems,
techniques  such as  the  use  of improved  operators and/or  the
diagonalization  of matrix  correlation functions \cite{Luscher:1990ck}
have been employed.
\footnote{For recent applications of these methods to multi-baryons, 
see e.g.  Ref.~\cite{Yamazaki:2011nd, Beane:2011iw}.}

Recently,  a novel  method to  derive hadron-hadron  interactions from
lattice    QCD    was    developed    by   HAL    QCD    collaboration
\cite{Ishii:2006ec}-\cite{Doi:2011gq}, where the
%   Nambu-Bethe-Salpeter (NBS) wave function at large $t$ is utilized to
Nambu-Bethe-Salpeter (NBS)  wave function  is utilized to  extract the
hadron-hadron potentials.
%%% 
% Since  it  relies  on  the  ground state  saturation,
Since its  extraction from hadronic correlation  functions relies on
the ground state saturation,
%%% 
the problems mentioned  above may exist in principle.   In this paper,
we  introduce  time dependent  Schr\"odinger-like  equation which  can
utilize the information  of moderate $t$ and can  avoid the problem of
ground state saturation.  This  allows one to derive the hadron-hadron
potentials  as  defined in  the  original  HAL  QCD method  with  less
systematic errors.  The key  observation is that the scattering states
with ``different" energies on the  lattice are governed by the ``same"
non-local   potential   $U({\vec  r},   {\vec   r}')$.   The   present
time-dependent  method has  already been  applied successfully  to the
baryon-baryon   ($BB$)   potentials    in   the   flavor-SU(3)   limit
\cite{Inoue:2010es,Inoue:2011ai}.   In the following,  we give  a full
theoretical account  of this imaginary-time  HAL QCD method  by taking
the  $NN$  scattering with  (2+1)-flavor  lattice  QCD  as a  concrete
example.

\section{Potential with Ground State Saturation}
\label{sec.review}

In the original HAL QCD method, the ``time-independent" NBS wave function
 was shown to satisfy the following  
 ``time-independent" Schr\"odinger equation \cite{Aoki:2009ji}:
\begin{equation}
  \left(
  \vec k^2/m_N - H_0
  \right)
  \psi_{\vec k}(\vec r)
  =
  \int d^3 r' U(\vec r, \vec r')
  \psi_{\vec k}(\vec r'),
  \label{eq.schrodinger}
\end{equation}
where $H_0 \equiv -\triangle/m_N$ with $m_N$ being the nucleon mass.
The potential $U(\vec r,\vec r')$ is non-local but independent on $k$
 \cite{Aoki:2009ji,KR:1956}.
The equal-time NBS wave function is given by
\begin{equation}
  \psi_{\vec k}(\vec x-\vec y)
  \equiv
%  Z^{-1}
  \langle 0 |  N(\vec x)N(\vec y)
  |N(\vec k)N(-\vec k); in\rangle,
\end{equation}
where  $|0 \rangle$  and $|N(\vec  k)N(-\vec k); in\rangle$  denote the
vacuum  and a  two-nucleon state  with an  asymptotic  momentum $\vec
k$, respectively, and $N(x)$ denotes a composite interpolating field for the nucleon.
%%%
For $N(x)$ being local,  the reduction formula for local composite operators
 \cite{HNZ}
 can be used to establish
the relation between the NBS wave function and the  S-matrix:
 The asymptotic
behavior    of     the NBS    wave    function     at    long    distance 
 reads
\cite{Aoki:2009ji, Lin:2001ek, Aoki:2005uf}
\begin{equation}
  \psi_{\vec k}(\vec r)
  =
  e^{i\delta(k)}
  \frac{\sin(kr + \delta(k))}{kr}
  +
  \cdots,
  \label{eq.asymptotic.behavior}
\end{equation}
where  $\delta(k)$ denotes  the (scattering)  phase of the S-matrix.
Therefore the $NN$ potential in Eq.(\ref{eq.schrodinger})
 gives correct phase shift $\delta(k)$ for all $k$
 in  the elastic  region  $E <  E_{\rm th}  \equiv 2m_{N}+m_{\pi}$, by construction.

 In lattice  QCD calculations, NBS  wave functions are  extracted from
 the $NN$ correlation function using the ground state saturation as
\begin{eqnarray}
 C_{NN}(\vec x-\vec y;t)
  & \equiv &
  \frac1{V}
  \sum_{\vec r}
  \langle 0|
  T[N(\vec x + \vec r,t) N(\vec y + \vec r, t)\cdot \bar{\mathcal{J}}(0)]
  |0\rangle
  \label{eq.four-point.correlator}
  \nonumber \\
  &=&
  \sum_{n}
  \psi_n(\vec x - \vec y)
  \cdot
  a_n e^{-E_n t} \nonumber \\
  &  \rightarrow &
  \psi_0(\vec x - \vec y) a_0 e^{-E_0 t} \quad (t\rightarrow\infty ),
\end{eqnarray}
where $\bar{\mathcal{J}}(0)$ is a two-nucleon source located at $t=0$,
%%%
$V$ denotes the spatial volume,
%%%
$\psi_n(\vec  x  -  \vec  y)   \equiv  \langle  0|  N(\vec  x)  N(\vec
y)|n\rangle$ denotes  an NBS wave  function for an  intermediate state
$|n\rangle$   with  the   energy  $E_n$,   and  $a_n   \equiv  \langle
n|\bar{\mathcal{J}}(0)|0\rangle$.
%%%
%%%
 In  the  spin-singlet  sector,  for example,  the  central  potential
 $V_{\rm  C}(r)$ in  the leading  order of  the velocity  expansion, $
 U(\vec r,\vec r') = V(\vec r,\vec\nabla_r) \delta^3(\vec r - \vec r')
 = \left\{  V_{\rm C}(r)  + O(\nabla^2) \right\}  \delta^3(\vec r-\vec
 r'), $ is given by
\begin{equation}
  V_{\rm C}(r)
  =
  \frac{\vec k^2}{m_N}
  -
  \lim_{t\to\infty}
  \frac{H_0 C_{NN}(\vec r,t)}{C_{NN}(\vec r,t)},
  \label{eq.prev.pot}
\end{equation}
where   $\vec  k$   denotes  the   ``asymptotic  momentum''   for  the
ground-state. The  ground state saturation is crucial  here to extract
the potential.

\section{Potential without  Ground State Saturation}
\label{sec.new_formalism}

In this section, we propose an alternative derivation of potential without using 
the ground state saturation.
For  this  purpose,  we   consider  the  normalized  $NN$ correlation function
\begin{equation}
  R(t, \vec r)
  \equiv
  C_{NN}(t, \vec r)/(e^{-m_N t})^2.
  \label{eq.r-correlator}
\end{equation}
We  here assume that $t$ is moderately large such that elastic
contributions    (at $E    <    E_{\rm   th}=2m_{N}+m_{\pi}$)    dominate
$C_{NN}(t,\vec r)$.

As before, we write
\begin{equation}
  R(t, \vec r)
  =
  \sum_{\vec k}
  \psi_{\vec k}(\vec r)
  \cdot
  a_{\vec k}
  \exp\left(-t \Delta W(\vec k) \right),
  \label{eq.spectral_decomposition}
\end{equation}
where $\Delta  W(\vec k)\equiv  2\sqrt{m_N^2 + \vec  k^2} -  2m_N$ and
$a_{\vec        k}\equiv        \langle        N(\vec        k)N(-\vec
k); in|\mathcal{J}(0)|0\rangle$. From an identity
$
  \Delta  W(\vec k)
  = 
  \frac{\vec k^2}{m_N} -  \frac{\Delta W(\vec k)^2}{4 m_N},
$
it is easy to see
\begin{eqnarray}
  - \frac{\partial}{\partial t}
  R(t, \vec r)
  &=&
  \sum_{\vec k}
  \left\{ \frac{\vec k^2}{m_N} - \frac{\Delta W(\vec k)^2}{4m_N} \right\}
  \psi_{\vec k}(\vec r)
  \cdot
  a_{\vec k}
  \exp\left(-t \Delta W(\vec k)\right),
  \label{eq.tmp}
  \\\nonumber
  &=&
  \sum_{\vec k}
  \left\{ H_0 + U - \frac1{4m_N}\frac{\partial^2}{\partial t^2} \right\}
  \psi_{\vec k}(\vec r)
  \cdot
  a_{\vec k}
  \exp\left(-t \Delta W(\vec k)\right), 
 \end{eqnarray}
where $U$ is the  integration kernel associated with the non-local potential
$U(\vec r,\vec r')$. Thanks to \Eq{eq.schrodinger}, 
$\vec k^2/m_N$ in the first line can be replaced  by $H_0 + U$ in 
the  second line.
%%%
We then arrive  at  the ``time-dependent''  Schr\"odinger-like
equation,
\begin{equation}
  \left\{
  \frac1{4m_N}\frac{\partial^2}{\partial t^2} 
  -\frac{\partial}{\partial t}
  - H_0
  \right\}
  R(t, \vec r)
  =
  \int d^3 r'
  U(\vec r, \vec r')
  R(t, \vec r'),
  \label{eq.tdep}
\end{equation}
which shows that the same potential $U(\vec r,\vec r')$ as defined
 in  \Eq{eq.schrodinger} can be obtained from $R(t,\vec r)$. 
An advantage  of this  method is  that 
the ground state saturation (or more generally  a single state saturation)
is not required for $R(t,\vec r)$  to satisfy \Eq{eq.tdep}.

Several comments are in order here:\\
(i) For  the present method to work,  $t$ has to be  large enough such
that  elastic contributions at  $E <  E_{\rm th}  = 2m_{N}  + m_{\pi}$
dominate $R(t,\vec r)$.  Note that such $t$ is  much smaller than that
required for the ground state saturation.
%%% 
This is especially so for the large volume,
%%%
since  typical size  of  gaps between  energy  eigenvalues shrinks  as
$O(1/L^2)$.  While it  becomes more and more difficult  to achieve the
ground state saturation for larger volume,
%%%
the  requirement of the  elastic dominance  $E <  E_{\rm th}$  is less
sensitive to the volume size.\\
(ii)  In the  leading order  of the  velocity  expansion, \Eq{eq.tdep}
leads to a generalization of \Eq{eq.prev.pot}
\begin{equation}
  V_{\rm C}(r)
  =
  -
  \frac{H_0 R(t,\vec r)}{R(t,\vec r)}
  -
  \frac{(\partial/\partial t) R(t,\vec r)}{R(t,\vec r)}
  +
  \frac1{4m_N}
  \frac{(\partial/\partial t)^2 R(t,\vec r)}{R(t,\vec r)}
  \label{eq.new-formula}
\end{equation}
for  the  spin-singlet   sector.
We  can  also   include  the  higher  order  terms  of the 
 velocity expansion as  discussed  in
\Ref{Aoki:2009ji,Murano:2011nz}.
Convergence   of   the   velocity   expansion in the original HAL QCD method
 can be   examined  
  by comparing local  potentials at  two different
energies as discussed in \Ref{Murano:2011nz}.
%%%
Equivalently, in the present method, the  convergence can be examined by
comparing the local potentials obtained for different $t$'s. \\
(iii)  $D   \equiv  \frac1{4m_N}  \frac{\partial^2}{\partial   t^2}  -
\frac{\partial}{\partial t}$  in \Eq{eq.tdep} 
plays a role of $\vec
k^2/m_N$ in \Eq{eq.schrodinger}.
For  $\Delta E \cdot t\gg  1$  where   the  ground  state  saturation  is  achieved,
\Eq{eq.tdep}  reduces  to \Eq{eq.schrodinger}.    We therefore can regard the
``time-dependent'' Schr\"odinger-like equation as an extension of
the time-independent  Schr\"odinger equation (\Eq{eq.schrodinger}). \\
(iv) In the non-relativistic limit
where 
$\Delta  W(\vec k)\equiv  2\sqrt{m_N^2 + \vec  k^2} -  2m_N
\simeq \frac{\vec{k^2}}{m_N}$,
``time-dependent'' Schr\"odinger-like equation leads to
\begin{equation}
  \left\{
%  \frac1{4m_N}\frac{\partial^2}{\partial t^2} 
  -\frac{\partial}{\partial t}
  - H_0
  \right\}
  R(t, \vec r)
  =
  \int d^3 r'
  U(\vec r, \vec r')
  R(t, \vec r') .
  \label{eq.tdep-nonrela}
\end{equation}
Therefore, the 2nd derivative term of $t$ 
%$\frac{1}{4m_N} \frac{\partial^2}{\partial t^2}$ 
in Eq.~(\ref{eq.tdep})
corresponds to the relativistic effect.
%If the ground state saturation is also achieved,
%we have $-\frac{\partial}{\partial t} \rightarrow E$
%and thus obtain an equation used in our previous studies%
%~\cite{Ishii:2006ec,Nemura:2008sp,Aoki:2009ji,Ishii:2010th,Inoue:2010hs, Murano:2011nz, Doi:2011gq}

%\end{enumerate}

\section{Numerical Results}
\label{sec.numerical_results}

To  test  the  present   method,  we  employ  (2+1)-flavor  QCD  gauge
configurations  generated by PACS-CS  collaboration \cite{Aoki:2008sm}
on $32^3\times 64$  lattice with the RG improved  Iwasaki gauge action
at  $\beta = 1.9$  and the  non-perturbatively $O(a)$  improved Wilson
quark  action at $(\kappa_{ud},\kappa_{s})  = (0.13700,  0.13640)$ and
$C_{\rm SW} =  1.715$.  This parameter set corresponds  to the lattice
spacing  $a\simeq  0.091$  fm  ($a^{-1}=2.176(31)$ GeV),  the  spatial
extent $L =  32a \simeq 2.90$ fm, $m_{\pi} \simeq  701$ MeV and $m_{N}
\simeq 1583$ MeV. %%%

The periodic boundary  condition is used for spatial directions, while
the Dirichlet  boundary condition is taken for  the temporal direction
% at  $t_{\rm DBC}=32  a$ ($-32  a <  t \le  32 a$),
at  $t_{\rm DBC}=32  a$ and $-32 a$,
%%%
to avoid opposite propagations  of two nucleons in temporal direction,
i.e, one  propagates forward and the other  propagates backward.  From
time-reversal and  charge conjugation symmetries, we  can average over
forward propagation at $t>0$ and backward propagation at $t< 0$ with a
source at $t=0$.
%%%
%%%
By temporally shifting gauge configurations, 21 source points are used
per  one configuration and  390 gauge  configurations are  employed in
total.  Statistical errors are  estimated by the Jackknife method with
a bin size  of 10 configurations.  Composite operators  for the proton
and the  neutron are taken to  be $ p(x)  \equiv \epsilon_{abc} \left(
  u^T_a(x)  C\gamma_5  d_b(x)  \right)  u_c(x)  $ and  $  n(x)  \equiv
\epsilon_{abc}  \left(  u^T_a(x) C\gamma_5  d_b(x)  \right) d_c(x),  $
respectively.  In our actual  calculation, we replace $e^{-m_{N}t}$ in
\Eq{eq.r-correlator} by the single-nucleon
%%%
%%%correlation
correlator
%%%
$C_{N}(t) \equiv \sum_{\vec x}\langle 0|T[N(x)\bar{N}(0)]|0\rangle$ to
suppress  statistical noises  of $R(t,\vec  r)$.  This  replacement is
permitted as  long as  the ground state  saturation for  $C_{N}(t)$ is
achieved.  Time derivatives in \Eq{eq.new-formula} are evaluated after
applying the polynomial interpolation of degree 5 to $R(t,\vec r)$.

In  order to see how excited  states of the two nucleons affect the 
final $NN$ potential, 
we introduce a source function with a real parameter $\alpha$ as
\begin{equation}
  f(x,y,z)
  \equiv
  1
  +\alpha\left(
    \cos(2\pi x/L) + \cos(2\pi y/L) + \cos(2\pi z/L)
  \right),
\end{equation}
%%%
which reduces to the wall source at $\alpha = 0$.
%%%
Two-nucleon  source is then defined by  $ \bar{\mathcal{J}}(f)
\equiv  \bar{p}(f)  \cdot   \bar{n}(f),  $  where  $\bar{p}(f)  \equiv
\epsilon_{abc}  \left(  \bar{u}_a(f)  C\gamma_5  \bar{d}_b(f)  \right)
\bar{u}_c(f)   $   and   $\bar{n}(f)  \equiv   \epsilon_{abc}   \left(
  \bar{u}_a(f) C\gamma_5 \bar{d}_b(f) \right) \bar{d}_c(f) $ with
%%%
$\bar{u}(f)\equiv   \sum_{\vec  x}   \bar{u}(\vec  x)f(\vec   x)$  and
$\bar{d}(f) \equiv \sum_{\vec x} \bar{d}(\vec x)f(\vec x)$.

\begin{figure}[b]
  \begin{center}
    \includegraphics[angle=-90,width=0.48\textwidth]{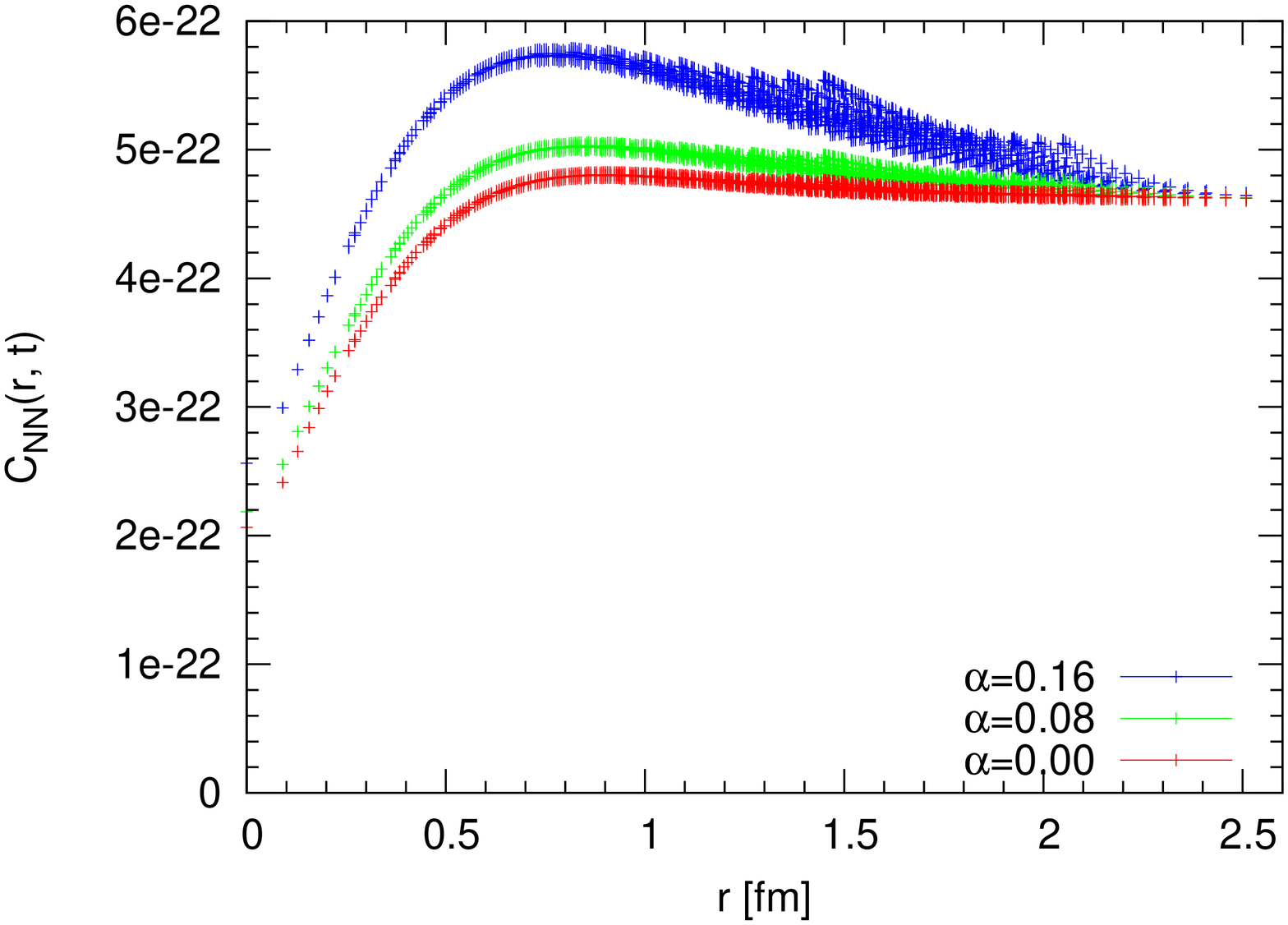}
    \includegraphics[angle=-90,width=0.48\textwidth]{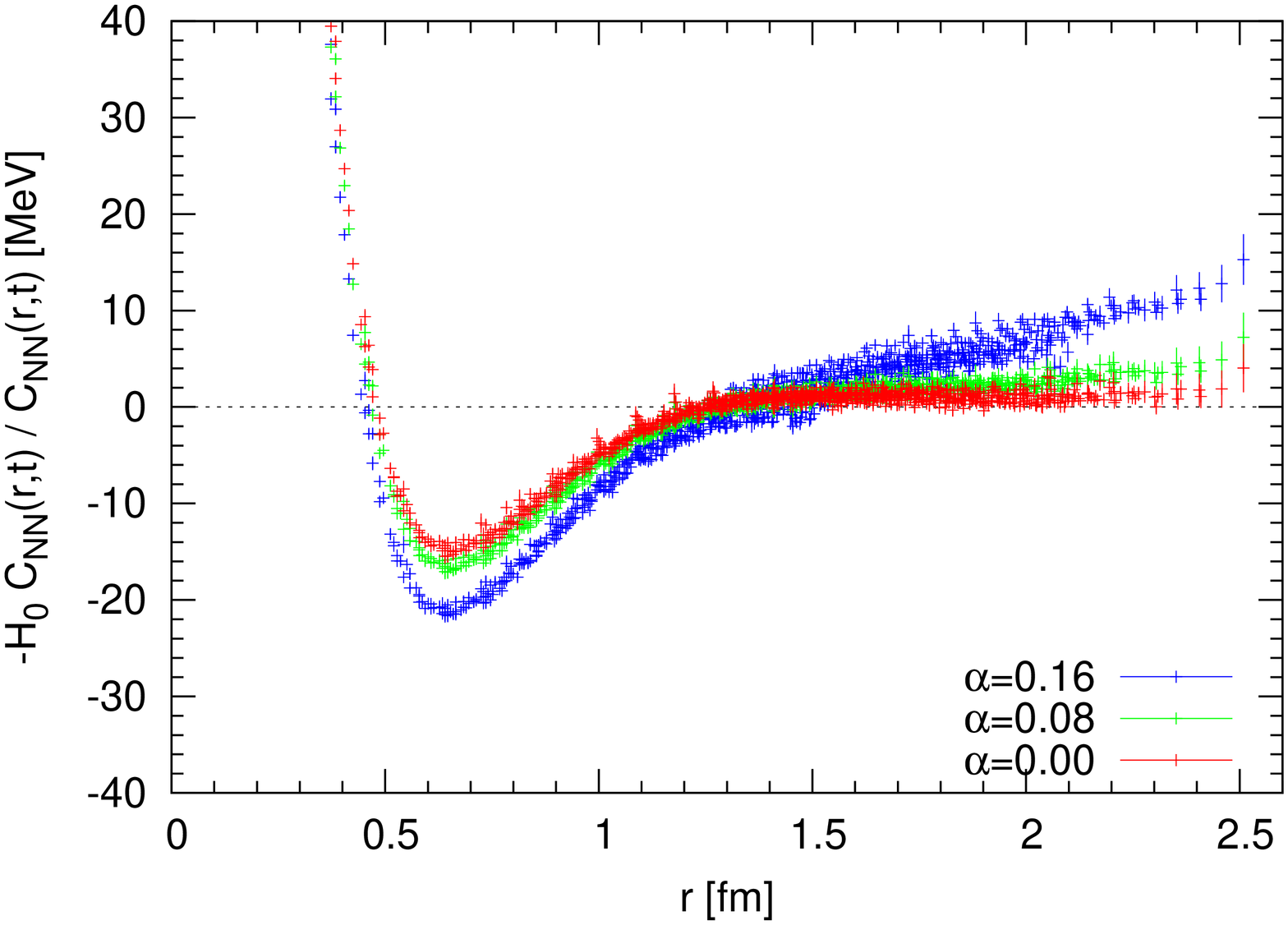}
  \end{center}
  \caption{(left)  $C_{NN}(\vec r,  t)$ at  $t =  9$.  (right)  $- H_0
    C_{NN}(\vec r, t)/C_{NN}(\vec r, t)$ at $t=9$.}
  \label{fig.1}
\end{figure}

\Fig{fig.1}  (left) shows $C_{NN}(\vec  r,t)$ at  $t=9$ for  $\alpha =
0.00$, $0.08$ and $0.16$.
%%%
%%%
If the  ground state saturation were achieved,  results with different
values   of   $\alpha$   should    be   the   same   up   to   overall
normalizations. \Fig{fig.1} (left) reveals that contamination from the
excited   states   is   non-negligible   at  $t=9$.    As   shown   in
\Fig{fig.1}(right), the  contamination is transferred  to the $\alpha$
dependence of $[H_0 C_{NN}(\vec r,t)]/C_{NN}(\vec r,t)$.
%%%

\begin{figure}
\begin{center}
  \includegraphics[angle=-90,width=0.48\textwidth]{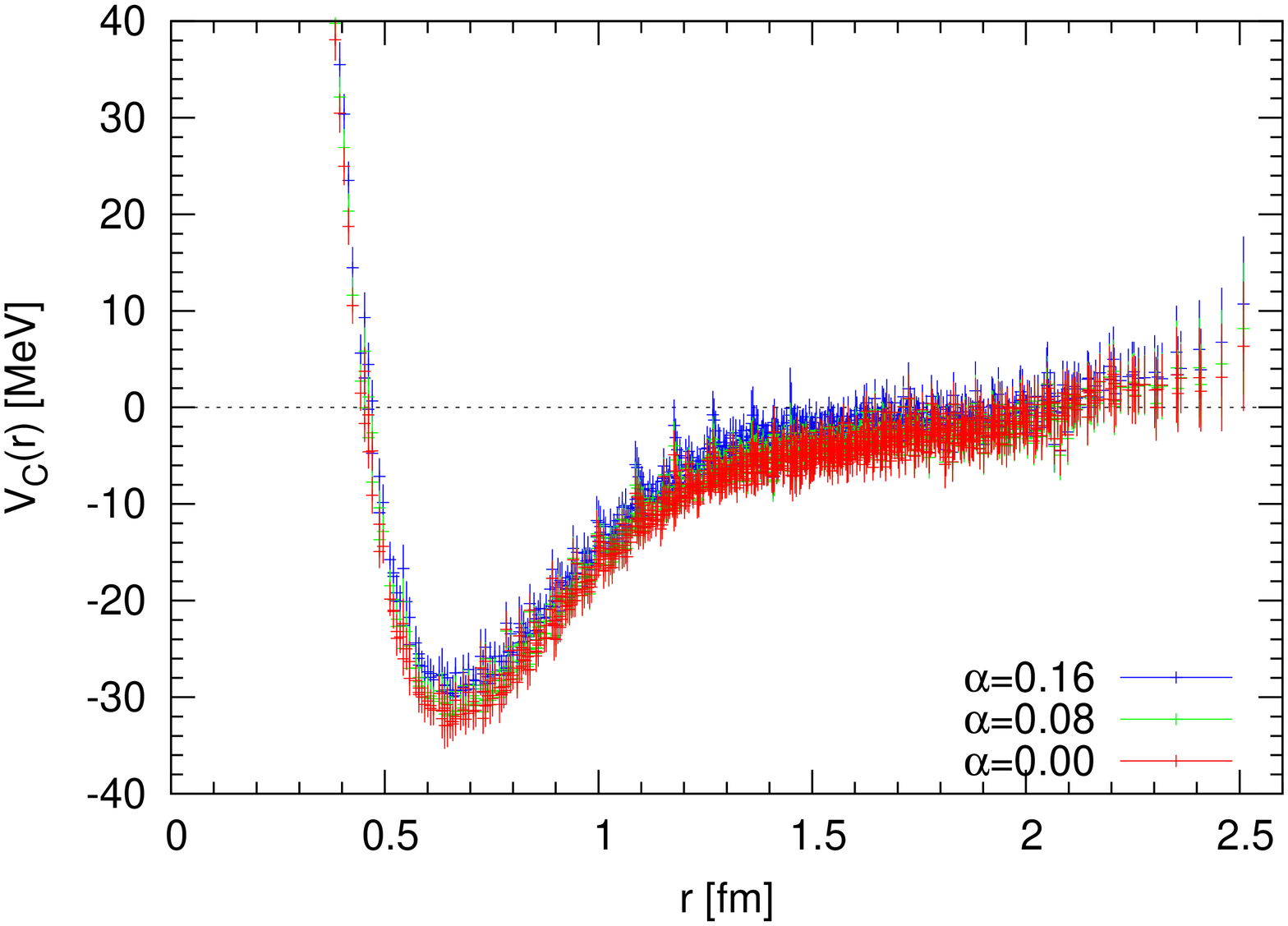}
  \includegraphics[angle=-90,width=0.48\textwidth]{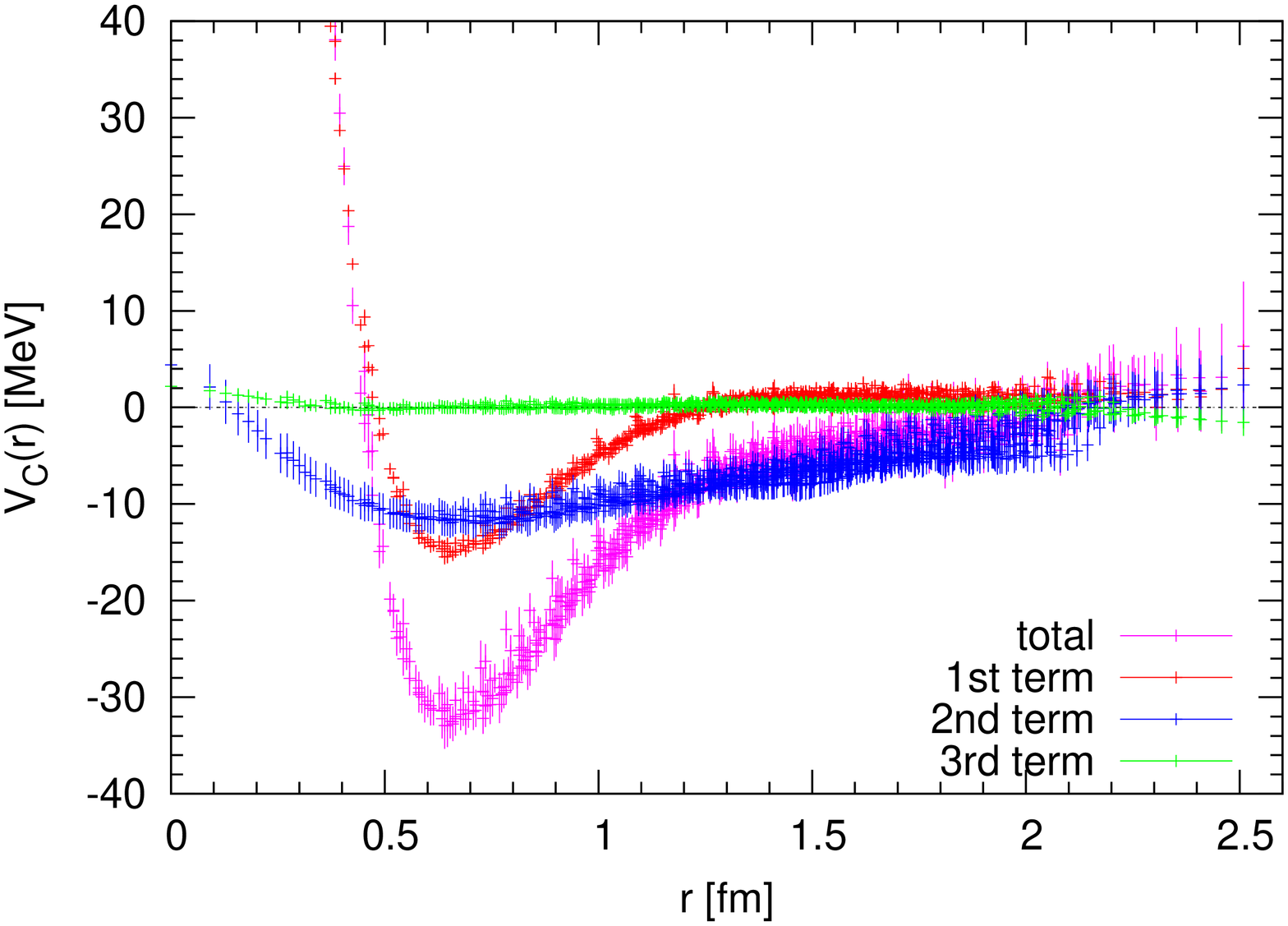}
\end{center}
\caption{(left)  Central  potentials   obtained  by  our  new  method
  \Eq{eq.new-formula} at $t=9$ for three values of $\alpha$.
  %%% 
  (right) Three contributions to $V_{\rm C}(r)$ in \Eq{eq.new-formula} at $t=9$ for $\alpha = 0$.
}
\label{fig.new-method}
\end{figure}

\Fig{fig.new-method}(left)  shows $V_{\rm C}(r)$  obtained from  our present 
method \Eq{eq.new-formula} for three values of $\alpha$.
%%%
%%%
The $\alpha$  dependence seen in  \Fig{fig.1}(right) disappears within
statistical errors.
%%%
%and $V_{\rm C}(r)$ is found to be more attractive.
%%%
Three   contributions to $V_{\rm C}(r)$  in
\Eq{eq.new-formula} are separately 
shown in \Fig{fig.new-method}(right)  for $\alpha = 0$.
%%%
We observe that the first  term of \Eq{eq.new-formula} (the red points)
determines the main trend, while the
second term (the blue points)
 gives an important correction.  The  third term (the green points),
  on the other hand,  is
negligible, showing  that  the  non-relativistic
approximation  $\Delta W(\vec k)  \simeq \vec  k^2/m_N$
works well in this case.
%%%
%Note that  the second term can  be regarded as  a point-wise effective
%mass defined by
%\begin{equation}
%  m_{\rm eff}(t; \vec x)
%  \equiv
%  \frac{\partial \log(R(t,\vec x))}{\partial t}
%  =
%  \frac{(\partial/\partial t) R(t,\vec x)}{R(t,\vec x)}.
%\end{equation}
Note   that  the   $\vec   r$-dependence  of   the   second  term   in
\Eq{eq.new-formula},     $-\frac{(\partial/\partial     t)    R(t,\vec
  r)}{R(t,\vec  r)}= -\frac{\partial \log(R(t,\vec  r))}{\partial t}$,
is a useful measure of the departure from the ground state saturation.

\section{Scattering Length and Scattering Phase Shift}
\label{sec.scattering_phase}

\begin{figure}
\begin{center}
  \includegraphics[angle=-90,width=0.48\textwidth]{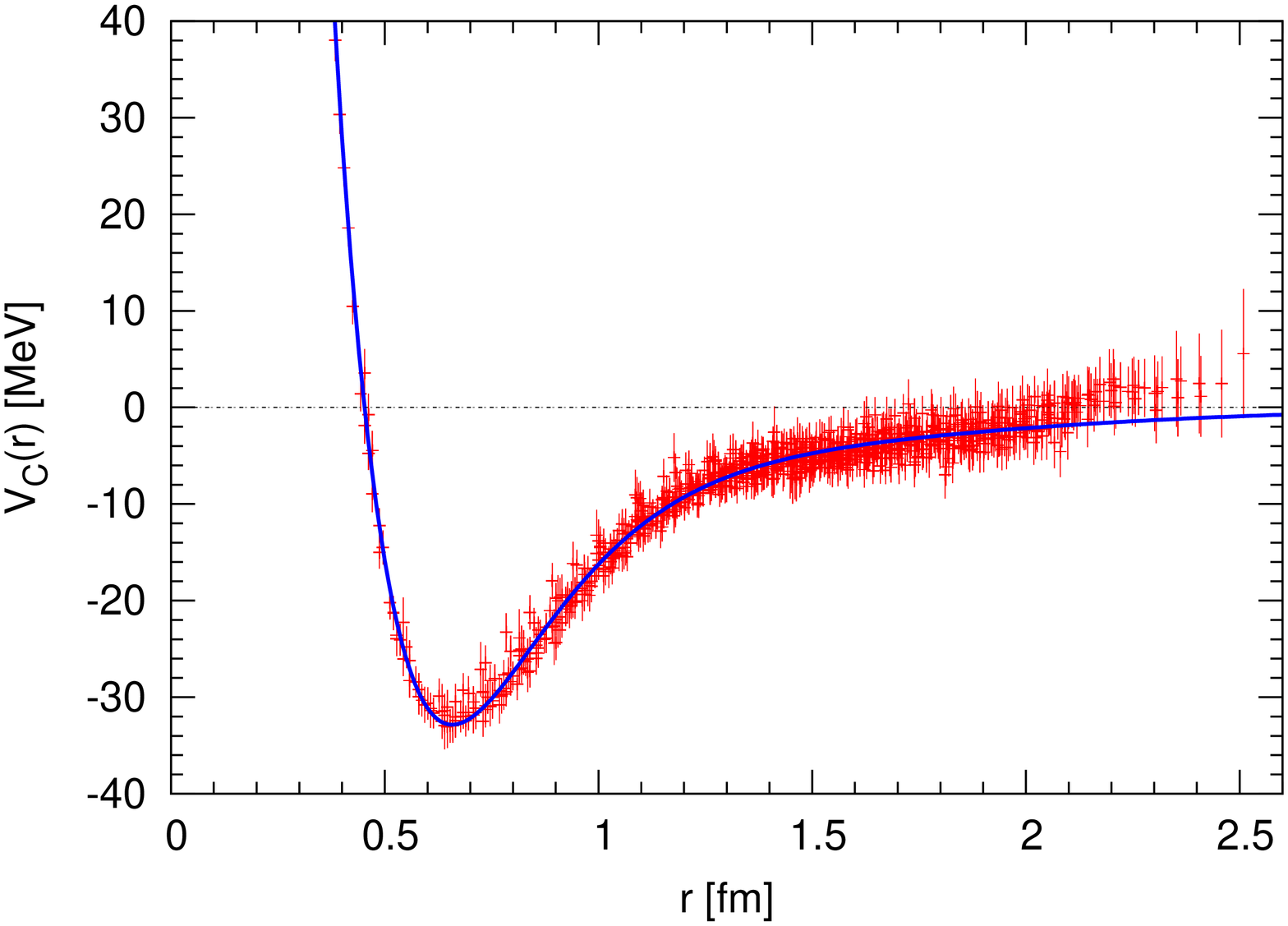}
  \includegraphics[angle=-90,width=0.48\textwidth]{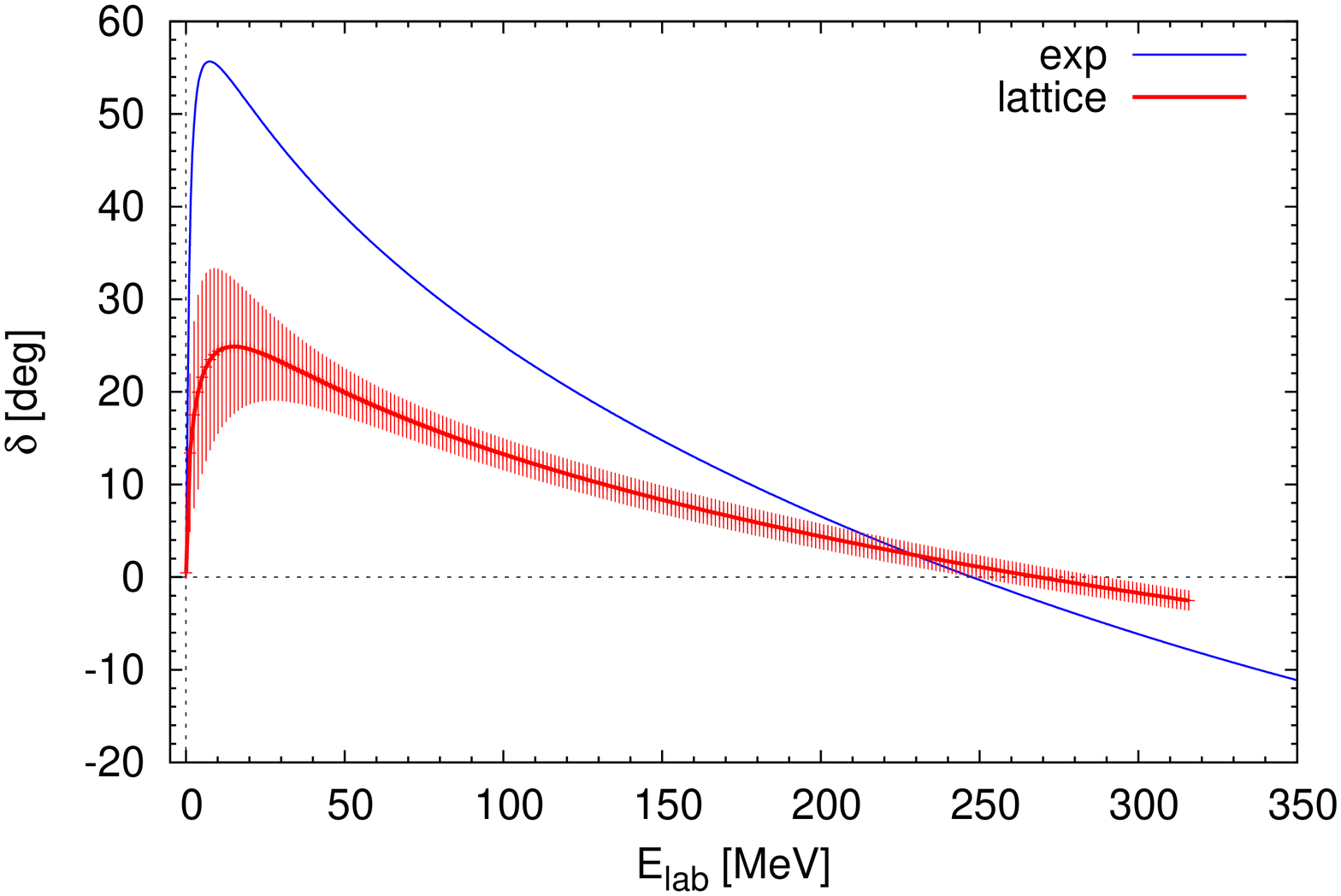}
\end{center}
\caption{(left)  The  multi-Gaussian fit  of  the  central
  potential $V_{\rm C}(r)$ with $N_{\rm  Gauss} = 5$ for $\alpha=0$ at
  $t=9$.   (right) The  scattering  phase in  $^1S_0$  channel in  the
  laboratory  frame obtained  from the  lattice NN  potential, together
  with experimental data\cite{nn-online}.}
\label{fig.scattering.phase}
\end{figure}
We now calculate the $NN$ scattering phase shift,
by solving the Schr\"odinger equation with the potential $V_{\rm C}(r)$ in the infinite volume.
For this purpose,  as shown in \Fig{fig.scattering.phase}(left), the central  potential $V_{\rm
  C}(r)$  is fitted with multi-Gaussian functions as
%%%
$
  g(r) 
  \equiv
  \sum_{n=1}^{N_{\rm gauss}}
  V_n \cdot \exp(-\nu_n r^2),
$
where  $V_n$ and  $\nu_n  (>0)$  are used  as  fit parameters,  $N_{\rm
  gauss}$  denotes the number of Gaussian functions.
We then solve the Schr\"odinger equation in  $^1S_0$ channel \cite{taylor}.
%by imposing the boundary condition, $\psi_{k}(r)
%\to j_0(kr)$ as $r \to 0$ \cite{taylor}.

\Fig{fig.scattering.phase}(right)    shows   the    scattering   phase
$\delta(k)$ extracted from the long distance behavior of the 
solution  $\psi_{k}(r)$, together  with the
experimental  data  for  comparison.   Qualitative feature of the phase shift  
as a function of $k$ 
is well reproduced, though the strength is weaker, most likely due to 
the heavy pion mass ($m_\pi \simeq 701$ MeV) in this calculation.
In fact, the recent 3-flavor QCD simulations
 show that the  $NN$ phase shift approaches toward the physical value  
as the quark mass  decreases \cite{Inoue:2011ai}.
%%%
The  scattering  length for $m_\pi \simeq 701$ MeV in the present method, 
calculated  from the  derivative  of  the
scattering phase  shift at $E_{\rm lab} =  0$, leads  to
 $a(^1S_0) =  \lim_{k\to 0} \tan \delta(k)/k = 1.6\pm 1.1$ fm.

\section{Non-local potential}

 By extending the present method further, one may directly extract
  the  non-local potential.   Let  us introduce  a  more general  $NN$
  correlation function,
\begin{eqnarray}
  R_{\vec r, \vec r'}(t)
  &=&
  {
    \frac1{V^2}\sum_{\vec x,\vec x'}
  }
  \frac{
    \langle 0 \vert
    T[ N(\vec x + \vec r, t) N(\vec x, t)
    \bar N(\vec x^\prime + \vec r^\prime, 0) \bar N(\vec x^\prime, 0)]
    \vert 0 \rangle}{(e^{-m_N t})^2},
\end{eqnarray}
which is shown to satisfy
\begin{eqnarray}
K( t) &\equiv& \left\{\frac{1}{4m_N}\frac{\partial^2}{\partial t^2} -\frac{\partial}{\partial t} - H_0\right\} R( t) 
= U\cdot R(t).
\label{eq:general}
\end{eqnarray}
Here the  matrix indices $\vec  r, \vec r^\prime$ for  $K(t)$, $R(t)$,
$U$ and a necessary integration over spatial coordinates are implicit.
The non-local potential  is then extracted as $  \tilde{U}= K(t) \cdot
\tilde R^{-1}  (t), $ where  an approximated inverse of  the hermitian
operator   $R(t)$   is   defined    by   $   \tilde   R^{-1}   (t)   =
\sum_{\lambda_n(t)\not=0}\lambda_n(t)^{-1} \vert n,t\rangle \langle n,
t  \vert .   $  Here $\lambda_n(t)$  and  $\vert n,t\rangle  $ are  an
eigenvalue  of   $R(t)$  and  its   eigenvector,  respectively.   Zero
eigenvalues  are  removed  in  the  summation.  Note  that  $U$  which
satisfies  \Eq{eq:general}  is  not  unique,  since $  U  =  \tilde{U}
+\sum_{\lambda_n(t)=0} c_n \vert n,t\rangle  \langle n, t \vert $ also
satisfies the same equation for arbitrary $\{ c_n\}$.  This is related
to the  fact that the  zero-modes or nearly zero-modes  are associated
with states above the inelastic threshold.

\section{Summary and Concluding Remarks }
\label{sec.conclusion}
A method  to extract hadron-hadron interactions by  generalizing the original
 HAL QCD method is proposed.
%%%
We  derived ``time-dependent''  Schr\"odinger-like equation,  a second
order differential equation in $t$,  which enables us to construct $NN$
potentials without assuming the ground state saturation in the elastic
region $E < E_{\rm th} = 2m_{N}+m_{\pi}$.
%%%
We have shown  that this method works well  for extracting the central
$NN$ potential in the $^1S_0$ channel: Identical potential is obtained
for  different source-operators within  the statistical  error.  Also,
resultant  $NN$   potential,  the  scattering  phase   shift  and  the
scattering   length   are  much   improved   both  qualitatively   and
quantitatively   from  those   obtained  by   assuming   ground  state
saturation.

While we have considered the system  only in the elastic region so far, an
extension   to   the   inelastic   region  is   also   possible.    In
Ref.~\cite{Aoki:2011gt},  it has been  shown that  one can  define and
extract the  hadronic potentials above  inelastic threshold.  Together
with the framework proposed in  this paper, this could provide a novel
prescription to solve the S-matrix in QCD.

\section*{Acknowledgments}
Lattice QCD  calculation has  been performed with  Blue Gene/L  at KEK
under the ``{\it Large scale simulation program}'' at KEK.
%%%
We thank PACS-CS Collaboration and  ILDG/JLDG for providing us the 2+1
flavor    gauge    configurations \cite{Aoki:2008sm,Beckett:2009cb,ildg/jldg}.
%%%
We are grateful for the  authors and maintainers of CPS++ \cite{cps}, a
modified version of which is used for measurement done in this work.
%%%
This  research is  supported in  part by  Grant-in-Aid  for Scientific
Research  on  Innovative  Areas(No.2004:20105001,  20105003)  and  for
Scientific   Research(C)  23540321,   JSPS  21$\cdot$5985   and  SPIRE
(Strategic Program for Innovative Research).



%% References


% \bibliography{<your-bib-database>}

\begin{thebibliography}{00}

\bibitem{Lepage:1989hd}
  G.~P.~Lepage,
%  ``THE ANALYSIS OF ALGORITHMS FOR LATTICE FIELD THEORY,''
  in {\it From Actions to Answers: Proceedings of the TASI 1989},
  edited by T.~Degrand and D.~Toussaint
  (World Scientific, Singapore, 1990).

\bibitem{Luscher:1990ck}
  M.~Luscher and U.~Wolff,
  %``HOW TO CALCULATE THE ELASTIC SCATTERING MATRIX IN TWO-DIMENSIONAL QUANTUM
  %FIELD THEORIES BY NUMERICAL SIMULATION,''
  Nucl.\ Phys.\  B {\bf 339} (1990) 222.
  %%CITATION = NUPHA,B339,222;%%

%\cite{Yamazaki:2011nd}
\bibitem{Yamazaki:2011nd}
  T.~Yamazaki, Y.~Kuramashi and A.~Ukawa,
  %``Two-Nucleon Bound States in Quenched Lattice QCD,''
  Phys.\ Rev.\  D {\bf 84} (2011) 054506
  [arXiv:1105.1418 [hep-lat]].
  %%CITATION = PHRVA,D84,054506;%%

%\cite{Beane:2011iw}
\bibitem{Beane:2011iw}
  S.~R.~Beane {\it et al.}  [NPLQCD Collaboration],
  %``The Deuteron and Exotic Two-Body Bound States from Lattice QCD,''
  arXiv:1109.2889 [hep-lat].
  %%CITATION = ARXIV:1109.2889;%%

  %\cite{Ishii:2006ec}
\bibitem{Ishii:2006ec}
  N.~Ishii, S.~Aoki and T.~Hatsuda,
  %``The Nuclear Force from Lattice QCD,''
  Phys.\ Rev.\ Lett.\  {\bf 99} (2007) 022001
  [arXiv:nucl-th/0611096].
  %%CITATION = PRLTA,99,022001;%%

  %\cite{Nemura:2008sp}
\bibitem{Nemura:2008sp}
  H.~Nemura, N.~Ishii, S.~Aoki and T.~Hatsuda,
  %``Hyperon-nucleon force from lattice QCD,''
  Phys.\ Lett.\  B {\bf 673} (2009) 136
  [arXiv:0806.1094 [nucl-th]].
  %%CITATION = PHLTA,B673,136;%%
  
  %\cite{Aoki:2009ji}
\bibitem{Aoki:2009ji}
  S.~Aoki, T.~Hatsuda and N.~Ishii,
  %``Theoretical Foundation of the Nuclear Force in QCD and its applications to
  %Central and Tensor Forces in Quenched Lattice QCD Simulations,''
  Prog.\ Theor.\ Phys.\  {\bf 123} (2010) 89
  [arXiv:0909.5585 [hep-lat]].
  %%CITATION = PTPKA,123,89;%%

%\cite{Ishii:2010th}
\bibitem{Ishii:2010th}
  N.~Ishii  [PACS-CS Collaboration and HAL-QCD Collaboration],
  %``Lattice study of nuclear forces,''
  PoS {\bf LAT2009} (2009) 019
  [arXiv:1004.0405 [hep-lat]];
%   %%CITATION = POSCI,LAT2009,019;%%
% %\cite{Ishii:2011zza}
% \bibitem{Ishii:2011zza}
%  N.~Ishii,
  %``Lambda-nucleon and nucleon-nucleon interactions on the Lattice,''
  Few Body Syst.\  {\bf 49} (2011) 269.
  %%CITATION = FBSYE,49,269;%%

%\cite{Inoue:2010hs}
\bibitem{Inoue:2010hs}
  T.~Inoue {\it et al.}  [HAL QCD collaboration],
  %``Baryon-Baryon Interactions in the Flavor SU(3) Limit from Full QCD
  %Simulations on the Lattice,''
  Prog.\ Theor.\ Phys.\  {\bf 124} (2010) 591
  [arXiv:1007.3559 [hep-lat]].
  %%CITATION = PTPKA,124,591;%%

  %\cite{Murano:2011nz}
\bibitem{Murano:2011nz}
  K.~Murano, N.~Ishii, S.~Aoki and T.~Hatsuda,
  %``Nucleon-Nucleon Potential and its Non-locality in Lattice QCD,''
  Prog.\ Theor.\ Phys.\  {\bf 125} (2011) 1225
  [arXiv:1103.0619 [hep-lat]].
  %%CITATION = PTPKA,125,1225;%%

  %\cite{Doi:2011gq}
\bibitem{Doi:2011gq}
  T.~Doi {\it et al.} [HAL QCD Collaboration],
  %``Exploring Three-Nucleon Forces in Lattice QCD,''
  arXiv:1106.2276 [hep-lat].
  %%CITATION = ARXIV:1106.2276;%%
%%%
    
  %\cite{Inoue:2010es}
\bibitem{Inoue:2010es}
  T.~Inoue {\it et al.}  [HAL QCD Collaboration],
  %``Bound H-dibaryon in Flavor SU(3) Limit of Lattice QCD,''
  Phys.\ Rev.\ Lett.\  {\bf 106} (2011) 162002
  [arXiv:1012.5928 [hep-lat]].
  %%CITATION = PRLTA,106,162002;%%  

%\cite{Inoue:2011ai}
\bibitem{Inoue:2011ai} 
  T.~Inoue {\it et al.}  [HAL QCD Collaboration],
  %``Two-Baryon Potentials and H-Dibaryon from 3-flavor Lattice QCD Simulations,''
Nucl. Phys. A (2012) in press
    [arXiv:1112.5926 [hep-lat]].
  %%CITATION = ARXIV:1112.5926;%%

\bibitem{KR:1956}
W. Kr\'{o}likowski and J. Rzewuski, Nuovo Cim. {\bf 4} (1956) 1212.
  
\bibitem{HNZ}
R. Haag, Phys. Rev. {\bf 112} (1958) 669;
K. Nishijima, Phys. Rev. {\bf 111} (1958) 995;
W. Zimmermann, Nuovo Cim. {\bf 10} (1958) 597.

%\cite{Lin:2001ek}
\bibitem{Lin:2001ek}
  C.~J.~D.~Lin, G.~Martinelli, C.~T.~Sachrajda and M.~Testa,
  %``K --> pi pi decays in a finite volume,''
  Nucl.\ Phys.\  B {\bf 619} (2001) 467
  [arXiv:hep-lat/0104006].
  %%CITATION = NUPHA,B619,467;%%

 %\cite{Aoki:2005uf}
\bibitem{Aoki:2005uf}
  S.~Aoki {\it et al.}  [CP-PACS Collaboration],
  %``I=2 pion scattering length from two-pion wave functions,''
  Phys.\ Rev.\  D {\bf 71} (2005) 094504
  [arXiv:hep-lat/0503025].
  %%CITATION = PHRVA,D71,094504;%%

  %\cite{Aoki:2008sm}
\bibitem{Aoki:2008sm}
  S.~Aoki {\it et al.}  [PACS-CS Collaboration],
  %``2+1 Flavor Lattice QCD toward the Physical Point,''
  Phys.\ Rev.\  D {\bf 79} (2009) 034503
  [arXiv:0807.1661 [hep-lat]].
  %%CITATION = PHRVA,D79,034503;%%


  \bibitem{taylor}
  J.R.~Taylor,
  Scattering Theory, The Quantum Theory of Nonrelativistic Collisions,
  Dover, 2000.
\bibitem{nn-online}
  http://www.nn-online.org/

\bibitem{Aoki:2011gt}
  S.~Aoki {\it et al.}  [HAL QCD Collaboration],
  %``Extraction of Hadron Interactions above Inelastic Threshold in Lattice
  %QCD,''
  Proc. Jpn. Acad., Ser. B, {\bf 87} (2011) 509
  [arXiv:1106.2281 [hep-lat]].

%\cite{Beckett:2009cb}
\bibitem{Beckett:2009cb}
  M.~G.~Beckett, B.~Joo, C.~M.~Maynard, D.~Pleiter, O.~Tatebe and T.~Yoshie,
  %``Building the International Lattice Data Grid,''
  Comput.\ Phys.\ Commun.\  {\bf 182} (2011) 1208
  [arXiv:0910.1692 [hep-lat]].
  %%CITATION = CPHCB,182,1208;%%
\bibitem{ildg/jldg}
  http://www.lqcd.org/ildg, 
  http://www.jldg.org
\bibitem{cps}
  CPS++. http://www.qcdoc.phys.columbia.index.html
  (maintainer: Chulwoo Jung)
\end{thebibliography}

% %% Authors are advised to submit their bibtex database files. They are
% %% requested to list a bibtex style file in the manuscript if they do
% %% not want to use model1-num-names.bst.

% %% References without bibTeX database:

\end{document}